# Our Mathematical Universe?

Jeremy Butterfield: Trinity College Cambridge, CB2 1TQ, UK
15 June 2014

<u>Review of: Max Tegmark, *Our Mathematical Universe*, Allen Lane 2014, £ 25, pp. 425</u>

Abstract: This is a discussion of some themes in Max Tegmark's recent book, *Our Mathematical Universe*. It was written as a review for *Plus Magazine*, the online magazine of the UK's national mathematics education and outreach project, the Mathematics Millennium Project (http://plus.maths.org/content/). Since some of the discussion---about symmetry breaking, and Pythagoreanism in the philosophy of mathematics---went beyond reviewing Tegmark's book, the material was divided into three online articles. This version combines those three articles, and adds some other material, in particular a brief defence of quidditism about properties. It also adds some references, to other *Plus* articles as well as academic articles. But it retains the informal style of *Plus*.

## I. A bigger, stranger cosmos

The message of this book is that the cosmos is much bigger, and much stranger, than you might have thought. Indeed, it is bigger and stranger in *ways* you might not have thought about.

Max Tegmark argues that it is bigger and stranger in four main ways: each one building on the last. Each represents such a vast extension of what we usually call the 'cosmos' (or 'universe') that he suggests we should use the word 'multiverse' for these extensions. So he calls the first main way it is bigger---the first main way he wants to surprise and unsettle you---the 'Level I Multiverse'. Then the second main way is the 'Level II Multiverse'; and so on till we get to the 'Level IV Multiverse'.

So get ready for a rapid gallop through---not just space, time and spacetime, as described by Einstein's general relativity and modern cosmology---but also the mysteries of how to interpret superpositions in quantum physics, and something maybe even more mysterious: the nature of mathematics itself.

But fear not: Max Tegmark is not only a gifted and distinguished cosmologist. He writes clearly and wittily about all the mind-bending ideas which his book expounds: not just the established physics of relativity, cosmology and quantum physics, but also the controversial possibilities---those four main ways for the cosmos to be bigger and stranger than you would have thought---that he wants to convince you about. Besides, he combines all this science with personal anecdotes about his own life and times: from his growing up in Sweden, to his later career in the USA. Thus the book begins with a nearly fatal bicycle accident which happened when he was a teenager (and the various quantum mechanical ways it might have been fatal!); and throughout the book, there are various anecdotes, e.g. about meeting the amazing Princeton physicist, John Wheeler. In short: ideal for readers of Plus Magazine . . . So, here goes!

## I. The Level I Multiverse

Level I is rather like the idea that maybe, when you were born, you were one of two twins—but the other twin got separated from you at birth, to be brought up far away, so that you never had the chance to meet each other. Except that for Tegmark's Level I, the birth in question is not yours, but the universe's!

That is to say, the suggestion is: at the Big Bang, there could have been pieces of matter that went off in another direction than did all the matter that went on to form our galaxy---and moreover, in a different direction from all the matter that makes the universe we can now observe---or even, will *ever* observe. Could that be so?

Well, the Big Bang is a tough topic to understand in detail! But a framework for understanding it, called inflationary cosmology, answers: 'Yes, it is so---and there are not just a few 'lost twins' of bits of early matter, but countlessly many: each of them zooming off, never to be seen by us, and also never to be seen by almost all other such bits of matter---or rather, by whatever observers those bits of matter might eventually aggregate into'. So inflationary cosmology proposes 'separating the children from one another at birth' on a truly cosmic scale!

But how much should we trust inflationary cosmology? There is good news and bad news. The good news is that in the thirty years since the idea of inflation was first invented (in about 1980), it has had several theoretical and even observational successes. (Cf. for example the Plus Magazine article on the Planck satellite: http://plus.maths.org/content/what-planck-saw.) So much so, that it is now physicists' orthodox view of the very early universe. The bad news is that---as always---we should also listen to the counsel of caution, the voice of scepticism. It says: we must distinguish between established facts and speculations.

This is the first of several counsels of caution ('health warnings') that I will need to issue as as we proceed through Uncle Max's Levels. Tegmark himself of course writes in a style of enthusiastic advocacy. But to his credit, he also intermittently pauses to emphasize that his stances are controversial, and also to reply in some detail to possible objections, or indeed to actual critics. See for example, his reply to George Ellis on pp. 360-363.

I should also note that this controversy is widespread among very distinguished cosmologists. For example, one recent critique of inflation (and specifically of the recent BICEP2 experiment) by Paul Steinhardt concludes that 'the inflationary paradigm is fundamentally untestable, and hence scientifically meaningless' (Nature, vol 510, 5 June 2014, p. 9: cf. http://www.nature.com/polopoly_fs/1.15346!/menu/main/topColumns/topLeftColumn/pdf/510009a.pdf ).

Agreed: modern cosmology has established that all the matter in the universe we can now observe was once in a hot, dense fireball about the size of the solar system. It has done this by combining precision cosmological observations, with established physical theories: especially general relativity, and the standard model of particle physics, which has for forty years stood up very well to countless tests, for example at the LHC in CERN, Geneva. (Cf. for example Plus Magazine's articles on the Higgs boson: http://plus.maths.org/content/hooray-higgs-edit-0, and http://plus.maths.org/content/higgs.) So 'the standard model' means our quantum field theoretic account of how particles like electrons and quarks behave at energies we can access. Those energies were attained in the early universe when it was about the size of the solar system, at about $10^{-9}$ seconds after the Big Bang. So this consilience of cosmology and quantum field theory at high energies is undoubtedly a stupendous scientific achievement.

Besides, our theories describing that hot, dense fireball can be extrapolated back still further. But as the radius becomes smaller, and the temperatures, densities and energies become higher, we leave established facts and enter the realms of speculation. And for the epoch at which inflation is supposed to have happened, we are most certainly far beyond the realm of established physics. That is: we are far beyond the energies at which we know both general relativity, and the standard model of particle physics, to be accurate. Indeed, at the end of inflationary epoch (if there was one!), the radius of the entire present observable universe was about 1 metre---wow!

But let us be bold! Let us accept inflationary cosmology's answer to our question 'Could that be so?' Let us accept that there are countless 'lost twins', separated into vastly many diverse universes, and that observers in those universes will never observe each other directly. This is the Level I Multiverse.

Before we hit Level II, we should note that Uncle Max has another mind-bending suggestion. Some inflationary models suggest there are *infinitely* many such universes, now mutually un-detectable. If so, there seems nothing to stop any specific scenario which you believe happened---some Tyrannosaurus Rex running through some Jurassic landscape, Napoleon being defeated at Waterloo, or Uncle Max's nearly-fatal bike accident---being played out, in replica form, in another of these universes. Indeed, there seems nothing to stop the scenario being played out, in replica form, any number of times across the whole set of universes. Indeed: even infinitely many times.

And what goes for a scenario which you believe happened, goes just as well for scenarios which you believe did <u>not</u> happen. Imagine Napoleon winning at Waterloo. Or rather, imagine someone very like Napoleon, in intrinsic respects (the hat, the arm in the waistcoat ...) and in historical role (an artillery engineer who became a conquering general . . .),  winning at a battle very like Waterloo (a flat landscape, soldiers in red coats sounding like Englishmen . . .). Imagine all this being filled out in some totally specific way. Of course all the myriad details vastly outstrip your powers of description: even if, to tell your tale, you could live much longer than three score years and ten. But nevermind the undescribability of it. Just imagine any such totally specific way in which (to put it briefly!) 'Napoleon wins at Waterloo'. Then the point is: there seems nothing to stop that exact scenario being played out in another of these universes. And similarly for all the countless other variants of (again: to put it briefly!) 'Napoleon winning'; and also, of course for all the countless variants of 'his losing'.

This is giddy stuff, to be sure. But so far, the alternative universes we have envisaged have been modest in one crucial respect. Yes, they vary in myriad ways: but the variations concern particular matters of fact, localized in space and time. A dinosaur runs, or doesn't. A man loses a battle, or wins it. There is, so far, no variation in the physical laws that govern processes. That is: assume for a moment that general relativity gives the correct laws for gravity, and quantum mechanics gives the correct laws for chemical reactions. Then the point is: so far, we have not envisaged a different law of gravity governing the dinosaur's run, or a different chemistry governing the general's digestion of his lunch.

But in inflationary cosmology, many models suggest that the physical laws <u>do</u> vary across the set of universes. If that is right, what we call 'physical laws' or 'laws of nature'---the deepest and most general patterns displayed by the ways in which physical events play out in our universe---would be better-called 'bye-laws'. For think of how the laws enacted by a municipal, rather than national, government are called 'bye-laws'; ('bye', like 'burgh' and 'borough', derives from the old Anglo-Saxon word for a settlement or town). This panoply of universes, differing in their laws, as well as their particular matters of fact, Tegmark calls the 'Level II Multiverse' . . .

## II. The Level II Multiverse

One immediately asks: exactly what kinds of variation (and how much variation of each kind), are being suggested? Of course, it all depends on the model chosen . . .

Here we should bear in mind that inflationary cosmology is a framework of ideas and results, not a theory that is defined with all due mathematical rigour. So neither are its individual models---and there is a lot of room for debate about which models are best.

This is regrettable, of course. But the point does not apply only to inflationary cosmology. The standard model---indeed any of our quantum field theories of interacting systems---is also <u>not</u> defined with all mathematical rigour. Accordingly, many brilliant people, working in the more mathematical areas of theoretical physics, are working to provide more rigorous formulations of quantum field theories, and thus also of inflationary cosmology.

(In fact the problems of rigour are deep: they originate in the theories' answers to straightforward questions involving troubling infinities. I will return to these problems when discussing renormalization at the end of this Section. And you can read more about them in *Plus* magazine's quantum field theory articles, at:
http://plus.maths.org/content/brief-history-quantum-field-theory.
Cf. especially http://plus.maths.org/content/problem-infinity)

So let us bear this warning in mind, but nevertheless proceed---and ask: what kinds of variation are being suggested?

At first sight, there seem to be two different ways in which 'the laws of physics could be different'. I will present them in Sections IIA and IIB. But as we will see at the end of this Section, inflationary cosmology suggests that the first way in fact happens as a special case of the second way.

## IIA. Varying constants

The first way is the obvious one: just imagine altering the equations! A lot of attention has focused on what seem the most modest sorts of alteration. Namely, not replacing 'the whole shebang' (like chucking the equations of general relativity in the bin, in favour of Newton's theory of gravity or other modern rivals of general relativity), but altering the values of parameters occurring in the equations, while keeping the same form for the equations, including which quantities they mention. These parameters are often constants of nature. For example: imagine altering the speed of light, or the strength of gravitational attraction as encoded in Newton's constant in his universal law of gravitation, or the ratio of the strengths of the gravitational and electromagnetic forces.

In the last several decades, there has been a lot of speculation about what the world would be like under such alterations (or combinations of them). And these speculations have been very fruitful, for two reasons. First: since one is 'only' (!) altering the values of parameters, this means that calculating what the world would be like is much more tractable than it would be, if one also altered the form of the laws---for who knows what alternative laws one should focus on?

Second: it has turned out, in countless such calculations, that even rather small alterations of parameters' values from their actual values would make for a world that is incompatible with life as we know it. For example, the altered physics would be incompatible with the formation of stars and of planets, and thus of the complex carbon-based chemistry on which life depends.

Here, we meet the idea of 'anthropic reasoning'. Roughly speaking, this is the idea that we might take this result---that only a narrow window around the actual value of a constant of nature is compatible with life as we know it---to be the best explanation of why the value is what it is. But this is not the place to pursue this topic. (For more discussion, cf. pp. 143 and 302 of Tegmark's book; or many articles in Plus magazine, such as http://plus.maths.org/content/planets-universes-part-I; and http://plus.maths.org/content/why-are-we-here;
Cf. also Wilczek (2008); the definitive book-length treatment remains Barrow and Tipler (1986).)

For my purpose here---namely, to summarize Tegmark's four-level multiverse---all I need at the moment is the idea of varying the laws of physics by making such variations in the constants. This is the first of the two ways of varying the laws; and as I mentioned, it is the more obvious one.

## II.B. Varying laws by breaking a symmetry

The second way is best introduced by an analogy with a crystal forming, and how such a process exhibits 'breaking a symmetry'. When a crystal forms, its atoms line up in a regular lattice; for example a cubical lattice like in Escher's famous picture. Such a lattice can be tilted one way or another: the rods do not have to be vertical and horizontal, as they are in Escher's picture. In fact, in a real crystal the lattice is not tilted in the exact same way throughout the crystal. Instead, the crystal is made up of tiny patches, called domains. In any one domain, comprising perhaps millions of atoms, the lattice is tilted in the same way. But the lattices in different domains will have different tilts.

Agreed: all this is hardly surprising. Typically the crystal forms by cooling a sample of liquid. (Remember the blue copper sulphate crystals you made in chemistry class at school, competing for who had the gentle touch to grow the biggest one!). The atoms settling down into a lattice is an example of that ubiquitous phenomenon in physics and chemistry: a system settling down into its state of lowest energy. Ideally, the lattice would have the same tilt throughout the entirety of the sample. The entire sample would form one vast domain, and the energy would indeed be lowest since there would be no domain-walls, i.e. lines or surfaces within the crystal where the tilt changes, with an associated energy cost. But that is of course very much an ideal, i.e. an unrealistic, situation. In reality, there will be a vast number of tiny events scattered throughout the cooling liquid (e.g. the presence of a speck of dust, or a passing cosmic ray): each of which can result in a shift in the local orientation of the lattice.

To sum up: we expect the physics and chemistry of the crystal's atoms to determine the general shape of the lattice, e.g. is it cubical, as in the picture? But we expect the exact orientation to be a matter of sheer happenstance, influenced by local events; and so it is very likely to be different, in different parts of the crystal.

There is a buzz-word for the variety in the tilts, the different tilts in the different domains: symmetry-breaking. The idea is: the physics and chemistry of the crystal's atoms, and the overall cooling process, does not prefer one tilt over any others. These principles allow the lattice any orientation in space: they have full rotational symmetry. But a specific state of the system, i.e. a specific domain, breaks the symmetry: it picks an orientation---because of microscopic causes that we may never know (nor care!) about.

This scenario, of symmetric principles having a non-symmetric instance, generalizes. Namely: a law, or collection of laws, of physics can have a symmetry, while a solution to those laws lacks it---and again, not because of some mystery, but simply as a matter of sheer happenstance. Once you grasp this idea, you realize that examples are everywhere: they are two a penny.

For example, Newton's laws of mechanics and his law of gravitation are rotationally symmetric. They encode no preferred orientation. Formally speaking:  if you calculate how your equation expressing these laws is transformed by rotating the coordinate system around any axis by any chosen amount (e.g. around the axis passing through the Earth's North Pole, by a 33 degree angle), you will find that the equation remains exactly the same.

On the other hand, almost all the solutions of Newton's laws, i.e. the instantaneous states of bodies (particles, planets ...) obeying Newton's laws, are not rotationally symmetric. They are not transformed into themselves by a rotation about any axis, by any amount. That would require that the solution is spherically symmetric, like a perfectly symmetric sphere of dust.  In short: the solution breaks the symmetry enjoyed by the law. Of course, this is in no way a problem: it is exactly what one would expect---the instantaneous state of a collection of bodies is usually not symmetric, and the lack of symmetry is no doubt usually a matter of sheer happenstance, of historical contingency—of where the bodies happened to be placed and how they happened to be moving.

So much by way of explaining the idea of symmetry-breaking. The idea is very important, quite apart from Tegmark's Level II multiverse. Suitably developed, it is crucial in various parts of modern physics. But for present purposes, its importance is that it engenders a second way in which one can envisage 'altering the laws of physics'.

To see this, imagine you are a tiny homunculus living in the domain of a crystal with, say, a cubical lattice; and knowing nothing of the world beyond that domain. As it happens, common table salt has a cubic lattice. So imagine you live in (a domain of) a crystal of salt.

Confined as you are to your tiny environment, you might well be unaware that the tilt of the lattice in your domain is a matter of sheer happenstance. You might well imagine that the tilt of your domain is as much a matter of law, as is the cubical structure of the lattice.

Thus in the jargon of 'bye-laws' which I introduced at the end of Section I: the tilt of your domain is in fact a bye-law---merely a regularity of your local environment. But since you know nothing of the world beyond your domain, you might well mistake it for a universal law.

The second way of  'altering the laws of physics' is now clear. The suggestion from inflationary cosmology is: we might be like the homunculus! What we take to be the laws of physics might be merely bye-laws: with different bye-laws being true in some, indeed most, of the other universes of the multiverse, universes which we will never observe.

So in the analogy with the crystal, the universes of the multiverse are like the domains of the crystal. And more specifically: these bye-laws might involve symmetry-breaking. They might break a symmetry of some genuine law---a law that holds good across the multiverse---by incorporating some or other matter of sheer happenstance---a matter that played out differently in different universes.

Indeed, the analogy with the crystal can be pushed further: just as the crystal's domains settle into various different lower-energy states during a global process of cooling, the multiverse's universes settle into various different lower-energy states during a global process of cooling.

Besides, it turns out that in some models of inflationary cosmology, this second way of 'altering the laws of physics' incorporates the first: which (cf. Section IIA) involved changing the constants of nature. Here, two other great themes of modern physics enter our story. One is called 'renormalization'. But it also has a more vivid label: 'running constants'. The other has only one name, but a very familiar one: string theory.

Thus the first theme is that a constant of nature can be dependent on the energy involved in the process concerned. The jargon is that the constant 'runs' with the energy scale; so people talk about 'running constants'. Agreed, this certainly sounds like weasel-words, to avoid sounding self-contradictory by saying 'varying constants'! (Or maybe it sounds like statistics about your training for the marathon, collated by your athletics coach!)

But actually, it is not self-contradictory. In fact, it reflects a profound physical idea: the idea of renormalization. This is a big topic which returns us to quantum field theory's troubling infinities, which I mentioned at the start of this Section. I will not go further into it. But, again, you can read more about it in Plus magazine quantum field theory articles. Among those articles, how renormalization tames the troubling infinities is emphasized in:
> http://plus.maths.org/content/rise-qed
> http://plus.maths.org/content/quantum-pictures
> and
> http://plus.maths.org/content/going-flow-0

So if we accept that a constant of nature can be dependent on energy, and that the universe--- okay, the multiverse!---is cooling down after the Big Bang like an enormous---okay, ginormous!--- fluid which is slowly crystalizing, then it makes sense that the constants might also settle down, as the energy decreases, to rather different values in different domains of the multiverse: i.e. in different universes of the Level 1 multiverse.

It is one thing to say 'it makes sense that the constants might also settle down to different values in different domains'. It is another thing to have a specific theory saying that indeed they do! And here enters the second theme: string theory.

Again, we first need a 'health warning'. Like the theories previously mentioned--- quantum field theories, inflationary cosmology---string theory is not yet defined in a mathematically rigorous way. (Indeed, it is, broadly speaking, more ill-defined than quantum field theory: most string theorists agree that what we call 'string theory' is really a collection of insights into a theory we do not yet 'have', in the way we do quantum field theory.) But let us again proceed, bearing in mind that the voice of scepticism is again whispering in our ears.

String theory is of course complicated. (For some details, cf. the Plus article http://plus.maths.org/content/string-theory-newton-einstein-and-beyond). But for the purpose of expounding Tegmark's multiverse, all we need is the fact that string theory indeed does propose that familiar constants of nature, such as the ratio of the strengths of the gravitational and electromagnetic forces, could settle down to different values in different domains of the Level I multiverse. Roughly speaking, each domain is associated with a different lowest-energy state of string theory.

(Here we meet other bits of jargon. (1) These different lowest-energy states are called 'vacua'. This jargon is taken over from quantum field theory. To be honest, it is misleading, since 'vacuum' suggests there is nothing, i.e. no system: while in this usage, there is always a system--- the various fields and strings---and 'vacuum' denotes a state of lowest energy. (2) The complicated set of possible states, with the various different minima, is called 'the landscape'.)

Thus we see how the second way of 'altering the laws of physics'---symmetry-breaking and us being like the homunculus in a domain of a salt crystal--- incorporates the first, i.e. altering the constants of nature.

So much by way of convincing you that the multiverse involves <u>varying laws</u>. Welcome to Level II!

### III. The Level III Multiverse

I turn to Uncle Max's Level III. It concerns the interpretation of quantum theory: a controversial area which we have not yet had to tangle with, despite referring to quantum field theories.

But I will discuss it only briefly. For it concerns a topic which readers of <u>Plus</u> magazine will be very familiar with. Namely: the paradox of 'Schroedinger's cat', and the proposal to resolve it by adopting the Everettian interpretation of quantum theory, also known as the 'many worlds interpretation'---the very phrase clearly suggests a multiverse!

Recall that 'Schroedinger's cat' is really a parable for making vivid the measurement problem of quantum theory. (For details, cf. the <u>Plus</u> magazine article: http://plus.maths.org/content/schrodingers-equation-what-does-it-mean). The problem is: the lack of values for physical quantities, such as momentum or position or energy, that is characteristic of quantum theory's description of microscopic systems, should also infect the familiar macroscopic realm of tables, chairs etc.

For example: measuring the momentum of an electron, which is in a state that is not definite for momentum (called: 'a superposition of momentum eigenstates') should lead to the pointer of the apparatus having no definite position. Rather, it should be in a superposition of states that are definite for position (called 'position eigenstates'). But this is not what we see! Witness the success of classical physics in describing countless physical systems as having at all times definite values for all physical quantities.

Schroedinger makes this problem vivid by imagining that a cat is fatally poisoned, or not, according as a microscopic quantum system e.g. an electron has, or does not have, some prescribed value for some prescribed quantity (e.g. value '5 units' for momentum in the x direction). Thus Schroedinger describes what he calls 'an infernal device' in which the opening of a vial of poison registers a prescribed result '5 units' for a measurement of an electron in a superposition of momentum eigenstates. So the cat ends up in a superposition of being fatally poisoned, and not: in a superposition of being dead and alive. Yikes!

It is now nearly eighty years since Schroedinger (in 1935) wrote his 'cat paper'; and nearly ninety years since he and Heisenberg (in 1925-26) put quantum theory in its established form, with its idea of superpositions of eigenstates. Unsurprisingly, the measurement problem was immediately recognized and vigorously debated, both by the new theory's architects like Schroedinger and Heisenberg, and by other luminaries like Einstein. (So it was not a matter of the problem going unrecognized until the 'cat paper' ten years later.) What is more surprising is that to this day---nearly ninety years later---it remains controversial.

So again, we need a health warning: we are again entering very contentious territory. But here it suffices to say one of the proposed solutions yields Uncle Max's Level III multiverse. Namely, Everett's proposal (in 1957: so more than fifty years ago) that, in short, a quantum superposition represents, not a set of alternatives of which one is real, but an actual plurality of realities.

In a bit more detail: the universe as a whole has a quantum state, which always evolves according to the Schroedinger equation: which is continuous and indeed deterministic---so there is no discontinuous 'collapse of the wave function'. Thus the universe contains a plethora of Everettian 'worlds' (also called 'branches'), where each such 'world' is something like the familiar macroscopic realm, with all tables, chairs etc. in definite positions, with definite momenta. But the worlds differ among themselves about these positions and momenta. Of course, each of us experiences only a single definite macrorealm. So the idea is: each of us just happens to be in one world, rather than another one. Or better: there are many versions of each of us, each version happening to be in one world rather than another. So this is indeed a proposal for a multiverse. Welcome to Level III! (To read more about the Everettian worlds, cf. the Plus article: http://plus.maths.org/content/parallel-universes)

Notice that in some ways, the same topics and questions---even conundrums---will arise for it as arose for the Level I multiverse. In that discussion, we envisaged that there are---very far away, and never observable by us---all sorts of scenarios involving diverse macroscopic objects, some of which might well be uncannily similar---even utterly similar---to familiar scenarios, like Napoleon losing the battle of Waterloo. And we ended by envisaging that some of the scenarios might be similar enough for us to describe them as containing a person like Napoleon and a battle like Waterloo---but in which that person (faux Napoleon, as we might call him) wins his battle (which we might call faux Waterloo).

Similarly here. Just as Schroedinger envisaged that an electron having one value of momentum rather than another ('5 units rather than 6') could lead to a cat's being dead or alive: so also, Everett and his advocates envisage that the two or more realities represented by some microscopic quantum superposition might lead to a macroscopic course of events, a history, 'branching'---of course, unobservably for the agents within it---into two (or more) histories. For example, a history could branch into some in which Napoleon loses the battle, and some in which he wins.

And here again, we have to be careful about language. On this view, 'Napoleon' really refers, not to a single person or human body, but to each of the many 'descendants', in the various branches, of the original person or body. Indeed, the Everettian vision is that there were, no doubt, countless microscopic quantum superpositions well before the date of the battle, appropriately correlated with the values of macroscopic quantities, so that the years leading up to the date of the battle involved many branchings, and the 'one Napoleon' whom we at first envisaged starting the battle (of whom some descendants would win, and others would lose) was himself one of many descendants of earlier stages of a 'Napoleon-like career'.

So there are some clear similarities between the ideas of Level I and Level III. But notice that the argument for believing in Level III has nothing to do with cosmology. The measurement problem, and the various proposals for solving it, and the arguments back and forth between these proposals, are all concerned with local physics, not with cosmology.

This difference is clear in another way also, when we recall Level II with its variation in the constants and in the form of the laws. The arguments for believing in Level II concerned inflationary cosmology and string theory, not the measurement problem. And similarly on the other side: nothing in the various proposals for solving the measurement problem suggests variation in the constants, or in the form of the physical laws.

IV. The Level IV Multiverse

And so to Tegmark's fourth and last dizzying vision. Unlike the first three Levels, this proposal is independent of the details of physics. It is distinctively philosophical: it is a claim about how mathematics describes physical reality. It is also a radical claim: namely, that physical reality—not merely is described by mathematics---but, instead, is mathematics.

About this fourth Level, I propose to be more critical than about the three others. After all, philosophical claims are, broadly speaking, more controvertible than physical claims. This is no doubt largely because the sorts of evidence counting for and against them is less sharply defined than for physical claims. So most philosophers accept that we make the best progress in assessing philosophical claims---in so far as we ever make progress!---by subjecting them to rigorous scrutiny.

So I will first discuss the claim that physical reality is mathematics, in general terms: indeed, in partly historical terms. Then I will state and assess Tegmark's argument in favour of it. But as I just warned: I will reject the claim. Sorry, Uncle Max!

I will proceed in five stages, giving each stage a Subsection. First I will state the claim, adding a bit of history (Section IV.A). Then I will state my main reason for rejecting it: that reason turns on introducing the distinction between pure and applied mathematics (Section IV.B). To properly justify my rejection, I will then give some details about that distinction (Section IV.C). In Section IV.D, I will use this distinction to rebut Tegmark's own argument for the claim: he argues that it follows from something we almost all accept: viz. that there is an external reality completely independent of us humans. Finally, in Section IV.E, I add to my rebuttal by briefly discussing a philosophical doctrine about the nature of physical properties: a doctrine called 'quidditism'.

<u>IV.A. From Pythagoras to Galileo and Beyond</u>

Philosophers have long had a name for the claim that physical reality is mathematics. They call it 'Pythagoreanism'. This refers to Pythagoras, the ancient Greek (ca. 572 – 497 BC). For millennia, he has been regarded as a mystic mathematician who, among other accomplishments, proved the theorem that bears his name, advocated the idea of mathematical proof, and believed that, in some obscure way, the world was made of numbers, i.e. of mathematics---hence the 'ism'.

(In fact, we now know, thanks to modern scholarship, that Pythagoras did none of these things. As often happens, it is a case of the sober truth being duller than the myth. For a glimpse of the details, cf. the review by Burnyeat (2007), citing the scholarly work of Walter Burkett.)

But setting aside the details of the historical Pythagoras, there is no doubt that the history of science, and especially of physics, provides countless illustrations of the power of mathematical language to describe natural phenomena. Famously, Galileo himself---the founding father of the mathematical description of motion---envisaged describing many, perhaps all, phenomena in mathematical terms. Thus in *The Assayer* (1623), he wrote the following (saying 'philosophy' in roughly the sense of our words 'natural science' or 'physics'):

'Philosophy is written in this grand book — I mean the universe — which stands continually open to our gaze, but it cannot be understood unless one first learns to comprehend the language in which it is written. It is written in the language of mathematics, and its characters are triangles, circles, and other geometric figures, without which it is humanly impossible to understand a single word of it; without these, one is wandering about in a dark labyrinth.'

Of course, since Galileo's time the language of mathematics has developed enormously---in ways that even he, a genius, would have found unimaginable. After all, already by the mid-seventeenth century the calculus was generalizing vastly from Galileo's 'triangles, circles, and other geometric figures', so as to consider arbitrary curves and their slopes (derivatives) and the areas under them (integrals).

(More precisely: the 'arbitrary' curves need to satisfy conditions, such as being suitably smooth, i.e. differentiable, so that they indeed have a derivative---conditions that are subtle to state, and were hard to discover, as the long history of rigorizing the calculus attests.)

As I have stated them so far, these developments seem to support Pythagoreanism. If mathematics gives us so powerful a language for describing nature, why not conclude that nature is mathematics?

As I said, Tegmark endorses this claim. (In fact, this quotation from Galileo is one of the mottoes of his Chapter 10, where he develops and defends his Pythagoreanism.) But he also ties this claim to his over-arching theme of a multiverse. That is: we should believe in the equal reality of all the possible mathematical structures. Welcome to Uncle Max's Level IV!

<u>IV.B. Beware: 'mathematical structure' is ambiguous</u>

As I announced, my aim in the rest of this review is to make sense of this claim and criticize it. This will of course involve clarifying the idea of 'mathematical structure' and thus the idea of all possible mathematical structures. But I can already state the gist of what Tegmark believes, and of my critique of it.

Tegmark does not claim only the following:

(1) 'Nature---that is the multiverse, as described so far, with its levels I to III---<u>is</u> a vastly complicated mathematical structure'.

He also claims:

(2) There is a mathematical multiverse: all possible mathematical structures are equally real.

So, putting (1) and (2) together, he claims:

(3) The multiverse, as described so far, with its levels I to III, <u>is</u> just one of countlessly many, equally real, such structures.

To which I say:

(i): I am happy to concur with (1). Well, happy enough for present purposes. But I think we should take seriously all those health warnings---Beware: here be speculations!---in the three previous Sections.

(ii): I am happy to say (2): 'There is a mathematical multiverse: all possible mathematical structures are equally real.' But I mean by this something different from what Tegmark means: both different and less controversial . . . let me explain . . .

In a word, my message to Uncle Max is: <u>Distinguish!</u> That is: we must distinguish between <u>applied mathematics</u> (also called: theoretical physics) and <u>pure mathematics</u>. And once we make this distinction, there are three important conclusions. We see that:

(a) It is for <u>applied</u> mathematics that claim (1) is tenable. That is: it is tenable that the multiverse, as described so far, with its levels I to III is an applied mathematical structure. And as I said: I concur with this.

(b) It is for <u>pure</u> mathematics that I am happy to say (2): 'There is a mathematical multiverse: all possible mathematical structures are equally real.'

Here I should note: I say 'happy' to be vivid, and for rhetorical simplicity. I admit that this belief in a vast plethora of pure mathematical structures is controversial. It is controversial because most philosophers maintain that all our concepts, beliefs and knowledge come from our experience of the empirical world: our experience of nature in space and time. That seems right to me. But it is hard to reconcile with our having beliefs, and even knowledge, about pure mathematical structures: especially infinite ones, since they certainly seem <u>not</u> to be given to us in experience.

Incidentally, this belief in a plethora of pure mathematical structures is called 'Platonism'. Note that this is <u>not</u> 'Pythagoreanism': because the topic is <u>pure</u> mathematics, not the mathematical description of nature. Thus I call myself a 'reluctant Platonist': 'reluctant' because Platonism is hard to reconcile with all our knowledge coming from our empirical experience.

(c) Platonism about pure mathematics (whether reluctant or enthusiastic) has nothing to do with a physical multiverse. To put the point in terms of propositions (1) to (3) above: if (1) says nature is an <u>applied</u> mathematical structure, and (2) asserts a plethora of <u>pure</u> mathematical structures, we cannot infer (3), that nature is one of many equally real structures.

Indeed, some logic books classify various tempting fallacies of reasoning. One of them involves two premises of an argument having a word in common---but occurring with different meanings. Assuming that in each premise the word has the meaning appropriate there, the premises are true; or at least, can be allowed to be true, for the sake of the argument. And the argument <u>looks</u> valid, when you look at the pattern of the words. But the conclusion is false. This is called 'the fallacy of equivocation'. An example, with apologies to Jane Austen and lovers of *Pride and Prejudice*: 'Elizabeth Barrett is going to the ball. To play cricket, we need the ball. Therefore: Elizabeth Barrett is going to what we need." (Note, incidentally, that the equivocal word, the culprit for the deceiving pattern of words---here: 'the ball'---need not occur in the conclusion.)

      To sum up: in Tegmark's propositions (1) and (2), 'mathematical structure' is equivocal between applied and pure structures. That in itself is no error. The reader who is aware of the distinction naturally reads (1) and (2) with the appropriate meaning: just as in my example's two premises, we read 'the ball' with the appropriate meaning. But it <u>is</u> an error to infer from (1) and (2), read with their respectively appropriate meanings, to (3).  For (3) needs to be read with 'mathematical structure' having one meaning or the other---applied or pure. But on either reading, it doesn't follow from (1) and (2). (Again: when they are read with their appropriate meanings---meanings that make them true, or at least, tenable for the sake of the argument.)

      In short: beware the fallacy of equivocation!

The upshot of this discussion is clear. One can believe in the multiverse including all the Levels I to III. And one can believe this multiverse is a structure in the sense of applied mathematics. And one can be a Platonist about pure mathematics, i.e. believe in ever so many pure mathematical structures. One can believe all this <u>without</u> believing that the physical multiverse---that 'supreme' applied-mathematical structure!---<u>is</u> a pure mathematical structure.

So much by way of stating what I called 'the gist of what Tegmark believes, and of my critique of it'. I apologize: for a gist, it was inordinately long. So after a gist as long as that, you may not want to hear further details! But rest assured: it <u>was</u> the main message of my Section IV. Thus what follows is only for <u>afficionados</u> and enthusiasts. But in the spirit of friendly philosophical debate---let us proceed!

There is more to be said, in three ways. First, I should give more explanation of my distinction between applied and pure mathematics: this I do in Section IV.C. Second, Tegmark gives a specific argument for the claim I reject: that the physical multiverse is a (pure) mathematical structure. So I need to address this argument: I will do so in Section IV.D. This leads to my final task. For in the course of his argument, Tegmark does briefly register that 'mathematical structure' is ambiguous in the sort of way I have expounded. So in Section IV.E, I take this up: it will lead us to a philosophical debate about the nature of properties.

<u>IV.C. Applied vs. Pure Mathematics</u>
To spell out the distinction between applied and pure mathematics, I will describe each in turn.

Applied mathematics gives---or at least: aims to give!---true descriptions of empirical, in particular physical, phenomena that are located in space and time. (Here 'true' does not mean 'complete': our description can be true even though typically the phenomena have---and we know them to have---myriad details that our description omits.)

For example, I spill my glass of milk and it spreads across the table. Applied mathematics successfully describes how it flows, in terms of relevant physical quantities, such as the positions, velocities and densities of small volumes of the milk. So in this example, the neglect of myriad details is, in part, a matter of neglecting the atomic constitution of milk. Of course, this is done by modeling the milk as composed of volumes that are large enough to contain many atoms (and so, we hope, to be unaffected by atomic phenomena); but which are also small by our human standards, so that the milk seems to be continuous in its make-up.

For my purposes, the crucial feature of this example is the mention of <u>relevant physical quantities</u>: in the example, the mention of position, velocity and density. Of course, it is one of the great glories of physics since Galileo's day that it has introduced new quantities---sometimes very arcane ones---and refined old quantities---sometimes in very subtle ways; and has combined the new and the old in a collection of laws and methods that---though fallible, and indeed changing with the passing decades---has gone from one success to another, both in theoretical understanding and in empirical quantitative prediction.

As examples of introducing new quantities, we might take (in roughly historical order): momentum, kinetic energy, potential energy, electric charge, entropy, spin.   As examples of refining old quantities in a subtle way, we might take (again, in roughly historical order): the refinement of the concept of mass (and so density) between the times of Galileo and Newton; Einstein's critique of absolute simultaneity in the Newtonian notion of time; and his unification of mass and energy.

Without doubt, these and similar examples of the development of physical concepts are among the great glories of physics. And this is so, wholly irrespective of what laws (e.g. laws of motion) the concepts may enter into. Indeed, this is something that every student of physics knows. Grasping the ideas involved in quantities like energy, entropy or spin, is a struggle and an achievement---and usually much harder than remembering the general equations, i.e. the laws, which mention the quantities!

So to round off my sketch of physics (or applied mathematics):--- We have come along way from Galileo's 'triangles, circles, and other geometric figures'. In his day, it was indeed reasonable to hope that for its concepts, i.e. its quantities, physics could manage with just the concepts of geometry (as inherited from the Greeks: triangles, circles, areas, volumes . . .) and maybe a little more---such as notions of contact or impact, and mass and-or density. This was precisely the agenda of the seventeenth century corpuscularian philosophy (also called: mechanical philosophy) spearheaded by Galileo. But it was not to be. Nature's imagination outstrips ours! So as physics proceeded to examine successive new domains of phenomena, it has had to introduce a succession of new quantities (and has also had to refine old ones). It is these distinctive physical quantities (and of course, their values for the system described) that are mentioned by the symbols in the equations of physics.

So to sum up: nowadays, we should revise Galileo's saying. Namely, instead of saying 'Nature is a book written in the language of mathematics', nowadays we should say: 'Nature is a book written in the syntax of mathematics, but with the semantics of physics'. (This aphorism is due to Ernan McMullin in his excellent paper on Galilean idealization (1985, p. 252-253).)

I can now be more precise about what at the end of Section IV.B I called an 'applied-mathematical structure'. In short: it is an assignment to each of a collection of objects of their values for various relevant physical quantities. Or to say it with a bit more physics jargon, and in a bit more detail:--

    (i)        Physics tends to think of an assignment as a function, to call objects 'systems', and to think of such an assignment (function) varying with time, to represent values changing over time.

    (ii)      So the short statement above applies to a collection of systems at a time: such an assignment (function) is an 'instantaneous state'. But the collection of such states as time varies---i.e. a history of the collection of systems--- would of course also count as an applied-mathematical structure.

    (iii)    Comments (i) and (ii) apply to field theories, where we think of physical quantities applying to spatial or spacetime points: e.g. the electric of magnetic field at points of space. It is just that here the objects in question are the points.

    (iv)    Though I will not go into detail: these comments also apply to quantum theories, in particular quantum field theories. The idea is that an instantaneous quantum state is specified by the quantities for which it is an eigenstate, i.e. definite in value. (To be strictly true, this statement requires that we set aside a subtlety, called 'superselection'; but let that pass.)

So much for applied mathematics. What of pure mathematics? What did I mean when above I shouted to my Uncle Max: 'Distinguish!'

My point was this. As every student of mathematics learns, in the mid-nineteenth century there arose, for several reasons, the idea of mathematics as the investigation of arbitrary structures. In other words: people started to think of mathematics as the investigation of the consequences of arbitrary rules that a mathematician postulates as governing some domain of elements. These rules are abstract in the sense that only their structural behaviour counts; and the elements are abstract in the sense that nothing is assumed about their natures, or relations to one another, except that they obey the announced rules.

To clarify this development, it is worth pausing over the history. I say this happened 'for several reasons' because one can discern different motivations in different branches of mathematics. Three obvious ones are as follows.

(1) In geometry, the rise of consistent axiomatizations of non-Euclidean geometries forced mathematicians to distinguish between (a) the actual geometry of physical space---an empirical matter that is revealed by the behaviour of rulers and protractors ('Do the angles of this triangle add up to 180 degrees?')---and (b) the idea of a pure system of geometry---which could be consistent, and worthy of investigation, even if it does not describe physical space. (Again, you can find out more about non-Euclidean geometry in a <u>Plus</u> article. Go to: http://plus.maths.org/content/non-euclidean-geometry-and-indras-pearls)

(2) In algebra, the investigations of Hamilton and Grassmann into new kinds of magnitude (e.g. quaternions), with new rules governing their operations of addition, multiplication etc., liberated people in a similar way. An algebraic system could be defined by axioms, and consistent, and worthy of investigation, even if it violated the rules of addition etc. familiar from arithmetic. For a <u>Plus</u> article on quaternions, go to: http://plus.maths.org/content/curious-quaternions

(3) Two other factors---the rise of set theory and logic, and the concomitant increasing concern to make deduction utterly rigorous, by formalizing mathematical language---also contributed to this conception of pure mathematics as the investigation of structure: in principle, of <u>any</u> structure.
(Both factors were themselves prompted by various paradoxical arguments that could be formulated in non-rigorous mathematical language, e.g. about sets. For more details about these paradoxes, cf. the <u>Plus</u> articles:
http://plus.maths.org/content/mathematical-mysteries-barbers-paradox
http://plus.maths.org/content/cantor-and-cohen-infinite-investigators-part-i)
Indeed, set theory and the formal languages introduced by the study of logic provided the perfect framework for this modern conception of pure mathematics. For the idea of a set is so simple, almost contentless! You just gather any old things together in a set by a pair of curly brackets, and you keep going, forming sets of sets, and sets of sets of sets, as much as you wish. And you are then free to postulate any structure you like on your sets (e.g. any new rule of 'addition' you like), and to investigate your postulates' consequences---by stating your postulates and deductions in the formal languages supplied by logic.

(For more details about how this conception of pure mathematics arose, cf. Kline's magisterial history of two millennia of mathematics (1972, especially pp. 1036-1038). Finally: in calling this the 'modern conception' of mathematics, I chose my words deliberately! For Gray (2008) makes a good case that this conception is 'modernist' in a similar sense to that in which art became 'modern' at about the same time, i.e. the mid-nineteenth century.)

Thus we arrive at the modern idea of a 'pure mathematical structure'. In short: it consists of a collection of objects, equipped with some postulated properties and relations e.g. an operation of multiplication subject to certain rules. The objects, properties and relations are all abstract in the sense nothing is assumed about their natures, except the behaviour postulated by the rules.

In a bit more detail: as I mentioned above, set theory turned out to be an ideal <u>lingua franca</u> for articulating this idea. Thus the idea of 'a collection of objects, equipped with some properties and relations' was made precise as 'a set of objects, with each postulated property being a subset, each postulated binary relation being a set of ordered pairs etc.'. The power of gathering things together in curly brackets---and iterating!

I need not go into further detail about this abstract and structural conception of pure mathematics. What matters for us is the main contrast with applied mathematics, and its structures. Namely: <u>there is no mention of physical quantities</u>.

Agreed: physical enquiry, and more generally empirical enquiry, may well have played a crucial heuristic role in leading a pure mathematician to articulate a structure. Of course, the history of mathematics contains many examples of this. But the resulting pure mathematical structure makes no mention of the physical quantities that inspired it. Being heuristically important does not imply being part of the official content of what results.

I agree that this 'lack of content' of a pure mathematical structure is perfectly compatible with the idea that something 'contentful', e.g. something mentioning physical quantities, is an instance of a pure mathematical structure. For the relation of instantiation allows a great deal of 'loss of content'.

For example: Here I am with my myriad properties, of size, shape, mass, temperature etc. Here is my family with its myriad properties and relations, e.g. our total mass, one person being more massive than another, one person being between two others etc. All this is 'contentful'. But I am an instance of many abstract 'contentless' properties: being less than two in number, not being a set etc. And similarly, my family as a collection of objects is an instance of many abstract 'contentless' properties and relations: being more than two in number, being the base-set of a group of order 6 etc.

But on the other hand: this 'lack of content' of a pure mathematical structure is incompatible with something 'contentful', e.g. something mentioning physical quantities, being a pure mathematical structure. For however we choose to make 'content' precise (e.g in terms of physical quantities or in some other way): something contentful cannot be literally identical with something contentless.

To sum up: The 'is' of identity, e.g. in 'a = b', is NOT the 'is' of instantiation, e.g. in 'Max is tall'. If indeed a= b, then a and b have the very same content, if any (in whatever sense of 'content' you like). For there is only one entity: a, which is also called 'b'. But in the case of 'Max is tall', Max can have countless properties, as contentful as you like (in whatever sense), that are not 'picked up on' or encoded by, or part of the meaning of, the predicate 'is tall'.

With that discussion in hand, my critique of Tegmark's Pythagoreanism---his claim that nature is mathematics---is very simply stated. It all turns on the simple but crucial distinction between the 'is' of identity and the 'is' of instantiation.

Thus I am willing to concede to Tegmark:

(A): The physical multiverse instantiates a pure mathematical structure. Indeed, being a Platonist about pure mathematics, though not yet convinced of his multiverse's first three Levels (remember the health warnings!): I would happily agree with Tegmark to the conditional claim:

(A'): If there is a physical multiverse (with whichever combination of Tegmark's first three Levels), then it instantiates a pure mathematical structure.

But still I deny:

(B): The physical multiverse is a pure mathematical structure.

The contrast between (A) and (B), between 'instantiates' and 'is', might seem a philosopher's quibble. But it is not. For it is claim (B), not (A), that is needed to get from Platonism about pure mathematics to Pythagoreanism: from Platonism about pure mathematics to Tegmark's Level IV multiverse!

### IV.D. Tegmark's argument from external reality

But Uncle Max has a specific philosophical argument for his Pythagoreanism, his Level IV multiverse: which I have so far not considered. So I owe you, and him, an assessment of it.

You will not be surprised to hear that I think the argument fails. It is invalid in the sense that it has a true premise but a false conclusion. Indeed, it is invalid because of the points in Sections IVB and IVC: i.e. the ambiguity of 'mathematical structure' that arises from the distinction between applied and pure mathematics. But at the risk of flogging a dead horse, I should state the argument and say why I think it fails.

Tegmark's premise is what he calls the 'External Reality Hypothesis' (ERH).

(ERH): There exists an external reality completely independent of us humans.

His argument is that (ERH) implies what he calls the 'Mathematical Universe Hypothesis' (MUH). MUH is essentially (B) above. In his exact words:

(MUH): Our external physical reality is a mathematical structure.

When you add to (MUH), Platonism about pure mathematics, that all the possible pure mathematical structures exist—which I agree to---you indeed get Tegmark's Level IV multiverse. That is, you get: our external physical reality (the multiverse with Levels I to III taken as accepted) is one of a plethora of equally real mathematical structures.

So what is the argument that (ERH) implies (MUH)? The idea is that since external reality is completely independent of us humans, it must have a description which is utterly free of subjective ingredients: that is, utterly free of factors arising from biological facts about human cognition, or cultural facts, or facts about an individual human's psychology.

Tegmark has a vivid metaphorical name for these subjective ingredients. He calls them 'baggage': the word 'baggage' connoting the burden or error in our description of nature due to the biases from our biological, cultural or individual history. He has an even more vivid metaphor for the effort to strip out such subjective ingredients from our description of nature: the effort which science, especially physics, has historically made, and should continue to make, so as to overcome the biases. Thus he calls getting rid of the subjective ingredients 'reducing the baggage allowance' (Chapter 10, especially pp. 255-265: the argument also occurs in his earlier paper (2008, pp. 102-108).

It is a vivid metaphor: and in an age when we are ruining the planet with too much jet travel, it seems a worthy aim . . .

And indeed, I agree with Tegmark that science, especially physics, has historically made successive efforts to overcome various cognitive biases due to our subjective constitution (biological, cultural or individual). I also agree that to make progress in the future, we must expect physics to continue to make that effort. Besides, many other scientists, especially physicists, and philosophers, would agree with him.

(In particular, many historians of philosophy would agree that modern physical science from the seventeenth century onwards has successively removed various subjective ingredients from the scientific world-view. For example, cf. Williams' discussion of what he calls the 'absolute conception of reality' (1978, pp. 65-67, 245-249).)

More specifically, concerning Tegmark's argument: I also believe (ERH). This is often called in philosophy, 'realism'. As it is usually put: the realist maintains that reality is utterly independent of the human mind. I also am happy to agree with Tegmark that (ERH) implies that reality must have a description which is utterly free of subjective ingredients. So Max's premise is true.

But obviously, this does <u>not</u> imply his conclusion (MUH)---for the reasons we saw in Sections IV.B and IV.C. That is: because of the distinction between <u>instantiates</u> and <u>is</u>, which I emphasized in terms of the physical quantities.

Thus, in terms of my labels (A) and (B) at the end of Section IV.C, I reply to Tegmark as follows:

(i)     (ERH) and the fact that reality must have an utterly objective description implies (A): The physical multiverse <u>instantiates</u> a pure mathematical structure. Or remembering the previous Sections' health warnings, they imply the conditional claim,  (A'): If there is a physical multiverse, then it <u>instantiates</u> a pure mathematical structure. But:

(ii)    They do <u>not</u> imply (B): The physical multiverse <u>is</u> a pure mathematical structure. That is, they do not imply (MUH).

IV.E Is a property 'more than' its web of relations to other properties?
In this final subsection (strictly for <u>aficionados</u> and enthusiasts!), I will raise some more philosophical controversy about the distinction between <u>instantiates</u> and <u>is.</u>

To his credit, Tegmark <u>does</u> register the distinction, albeit briefly: namely at p 260, footnote. There is more discussion in a paper that preceded his book, where he writes (2008, p.107):

'Whereas the customary terminology in physics textbooks is that the external reality is <u>described by</u> mathematics, the (MUH) states that it <u>is</u> mathematics (more specfically a mathematical structure). This corresponds to the 'ontic' version of universal structural realism in the philosophical terminology of Ladyman (1998) and McCabe (2006)'.

I reply as follows, with three points, (a) to (c).
(a): I agree that some speculative philosophical positions may collapse the distinction, saying 'is' where the customary view says 'instantiates'. But:
(b): I agree that McCabe's discussion seems to endorse collapsing the distinction. But on my understanding of Ladyman's position—whose paper (1998) named and advocated 'ontic structural realism'---he does not wish to collapse the distinction. More important:
(c): Setting aside whatever might be the views of these authors, we should <u>not</u> collapse the distinction.

My reason for saying (c) is that to collapse the distinction, undoubtedly the best strategy is to say that a property or relation, or at least the properties and relations mentioned by physics---such as the physical quantities that I used in Section IV.C to ram home the distinction between <u>instantiates</u> and <u>is</u>---is completely specified by its web or pattern of relations to other properties and relations. The paradigm such "relation to other properties and relations" is the relation of being associated together in a law. For example, the quantities force and acceleration are related by Newton's second law.

And so the strategy in question maintains that everything there is to know about any physical quantity, e.g. the entire nature of force, can be recovered from the web or pattern of relations among quantities---each of which is thus specified only as a node in this web. In mathematical terms, this requires at least that the web is to have only the identity map as an automorphism. The idea is: if this is so, then the apparent 'contentfulness' (cf. the end of Section IV.C) of physical quantities can be 'analysed away' as a matter of structure. That is: we seem to glimpse how the applied mathematical structures, with their 'contentful' physical quantities could turn out to be, <u>au fond</u>, pure mathematical---'contentless', 'abstract', 'structural'.

But I say: this strategy is wrong (and so the distinction between <u>instantiates</u> and <u>is</u> does not collapse). For there is every reason to think that at least some physical quantities each have a nature that is <u>not</u> exhausted by their web or pattern of relations to other properties and relations.

As I admitted in (a), I agree that some philosophical positions deny this. There is even a jargon. To say, as I do, that at least some properties (or relations) have a nature that is <u>not</u> exhausted by their web of relations to others, is to be a <u>quidditist</u>. (This is from 'quid', the Latin for 'what': for the idea is that a property has a 'content' or 'whatness'.)  So in effect, I am saying that for Uncle Max to defend his argument that (ERH) implies (MUH), his most promising strategy is to deny quidditism. But my objection to him is: quidditism looks right! As the following example suggests.

(Agreed: my example of electric charge, and the flip between positive and negative values of charge, is subtler than my discussion will suggest. (For it leads in to the question of the proper treatment of time-reversal: for which cf. e.g. Malament (2004).) So I do not claim that the case of electric charge gives conclusive proof of quidditism. But I <u>do</u> claim that it makes quidditism plausible; and thus that it shows Tegmark should have worked harder to justify anti-quidditism---to replace <u>instantiates</u> by <u>is</u>.)

Take the two properties, having negative electric charge, and having positive electric charge. Or if you prefer some precise amount of each, e.g. having plus or minus one Coulomb of charge.

Imagine them (or one of them) occurring in an applied mathematical structure, i.e. a physically possible state of affairs, according to some theory. For example, let the theory be electromagnetic theory, and let the properties (quantities or exact values of them) occur in a possible instantaneous state for a collection of particles. Thus: here is a particle with charge +3 Coulombs, there is another with charge -2 Coulombs. (Recall the sketch of applied mathematical structures in (i) t o (iv), in Section IV.C.)

Now imagine flipping positive and negative charge. Here, I do not mean changing our linguistic convention about what amounts of charge we call 'negative' and which 'positive'. I mean an active transformation which changes the state of affairs: in the new state of affairs, here is a particle with charge -3 Coulombs, there is another with charge +2 Coulombs. (I am using the same language as before!)

I agree that this new state of affairs of affairs will behave, e.g. change over time, in a manner that mirrors the behaviour of the original state of affairs. Here, the 'mirroring' can be spelt out in detail using the idea of a <u>symmetry of the laws.</u> I discussed this in Section II.B. In terms of the current idea of the laws as a web of relations between quantities, a symmetry is a transformation of the web, i.e. a function sending nodes to nodes etc., that preserves the web's structure e.g. the connectivity relations between nodes. But I do not need to go into details about the mirroring of the original state of affairs' behaviour.

For the relevant point---the point I want to emphasize---is a different, and a simpler, one. It is just that the new state of affairs is <u>different</u> from the original.

And just that point is enough to imply that there are facts---'content' or 'nature'---about, say, negative charge, or having an exact value of negative charge (e.g. -2 Coulombs) that <u>outstrip</u> its web of relations to other quantities: that <u>cannot</u> be 'analysed away' in terms of the quantity's position as a node in the web. For the 'flip' being a symmetry of the laws implies that this web of relations <u>is</u> shared with positive charge (or the corresponding exact value: in my example, +2 Coulombs): but the states of affairs are different!

## Acknowledgments

I am very grateful to Marianne Freiberger, the editor at *Plus*, for the invitation to review the book, and for dividing and editing the bulk of this discussion into three separate articles appropriate for *Plus*. They can be found at these URLs:
The review: http://plus.maths.org/content/mathematical-universe-0
Discussion of symmetry breaking: http://plus.maths.org/content/breaking-symmetry
Discussion of the world being mathematics: http://plus.maths.org/content/world-made-maths
I am also very grateful to Harvey Brown, George Ellis, Huw Price, Joe Silk, David Sloan and David Wallace for comments on a previous version: I wish I could have incorporated all their wisdom!